%% file: absMF.tex
\documentclass[fleqn,usenatbib]{aastex701}

\usepackage[T1]{fontenc}
\usepackage{ae,aecompl}
\usepackage{newtxmath} 

\usepackage{graphicx}	
\usepackage{amsmath}	

\input mymacros_nojournals

\begin{document}

\title[Absorber Masses]{Mapping the $z\geq5$ SiIV Column Density Distribution onto the Galaxy Stellar Mass Function Using the Cumulative Absorption Cross Section}

\author[orcid=0000-0002-0496-1656]{Kristian Finlator}
\affiliation{New Mexico State University, Las Cruces, NM, USA}
\affiliation{Cosmic Dawn Center (DAWN), Niels Bohr Institute, University of Copenhagen / DTU-Space, Technical University of Denmark}
\email[show]{finlator@nmsu.edu}

\author[orcid=0009-0002-2449-2508]{Sam Patterson}
\affiliation{New Mexico State University, Las Cruces, NM, USA}
\email{samk@nmsu.edu}

\author[orcid=0009-0001-4440-2568]{Nora Nava}
\affiliation{New Mexico State University, Las Cruces, NM, USA}
\email{nnava5@nmsu.edu}

\author[orcid=0009-0004-6379-2460]{Ayanah Cason}
\affiliation{Vanderbilt University} 
\email{ayanah1121@gmail.com}

\author[orcid=0000-0002-0761-1985]{Samir Ku\v{s}mi\'{c}}
\affiliation{New Mexico State University, Las Cruces, NM, USA}
\email{samirkusmic@gmail.com}

\author[orcid=0009-0004-8503-0483]{Ezra Huscher}
\affiliation{New Mexico State University, Las Cruces, NM, USA}
\email{ez@nmsu.edu}

\author[orcid=0000-0002-0072-0281]{Farhanul Hasan}
\affiliation{Space Telescope Science Institute, 3700 San Martin Drive, Baltimore, MD 21218, USA}
\email{fhasan@stsci.edu}

\begin{abstract}
Efforts to constrain directly the activity in low-mass galaxies confront sensitivity limits even in the JWST era. Metal absorbers offer a complementary probe and are easier to detect, but leveraging them requires a known relationship between absorber strength and host mass. To this end, many studies assume a simple monotonic relationship between absorber strength and host mass. This \emph{ansatz} ignores evidence that galaxies at fixed luminosity host absorbers spanning a variety of strengths. We address this issue by deriving a six-parameter model for the cumulative absorption cross section from cosmological simulations that combines with the galaxy stellar mass function to predict the absorber column density distribution (CDD). A maximum-likelihood analysis confirms that this approach reconciles the observed galaxy stellar mass function with the observed SiIV CDD at $z=$5--6. The extrapolated CDD grows uncertain outside the observed range and the resulting constraints contain degeneracies, highlighting the need for improved measurements. Galaxies of all masses host absorbers of all strengths, but a weak empirical association between massive galaxies and strong absorbers is indicated. Faint galaxies ($M_* < 10^8\msun$) host the majority of weak SiIV absorbers ($\log(N/\mathrm{cm}^{-2}) < 13$), emphasizing emission/absorber complementarity. The assumption of a power-law relationship between absorbers' geometric cross sections and host galaxy masses is empirically disfavored. The model may be applied to any combination of ion and redshift if the galaxy stellar mass function is well-constrained. Future observational tests incorporating improved host statistics will extend the model’s range. 
\end{abstract}



\section{Introduction} \label{sec:intro}
The early abundance of low-mass galaxies remains a central uncertainty in models of early galaxy formation and cosmological reionization. Radiative cooling is expected to occur efficiently in halos down to the HI cooling limit ($\gtrsim10^8\msun$ at $z<8$) and perhaps less efficiently in molecularly cooled minihalos (for example,~\citealt{nebrin23,dhandha2025}). As soon as star formation commences, this cooling competes with negative kinetic and radiative feedback from forming stars, evolved massive stars, active nuclei, and ablation by outflows from neighboring galaxies~\citep{scanBroad2001}. These effects are, in turn, partially offset by positive feedback from metal-enhanced cooling by ions and dust. Following HI reionization and its associated heating, the minimum halo mass for efficient star formation rose to its present-day value of $10^{10}\msun$~\citep{thoul96,pw2023}. Owing to uncertainties associated with these individual processes as well as their nonlinear couplings, the actual dependence of stellar mass on halo mass at low masses $<10^{10}\msun$ remains uncertain. For example, theoretical models predict that the UV luminosity function in the EoR flattens at a luminosity ranging variously from -8 to -14 (cf.\ Section 4.4 of~\citealt{bouwens2022}).

Observational exploration of these ideas requires probes that are uniquely sensitive to early, low-mass galaxies. Recent observations suggest that the galaxy rest-frame ultraviolet luminosity function rises steadily to luminosities $M_\mathrm{1500}=-15$ in the EoR~\citep{fink15a,bouwens2022,atek2024,harikane2024} and to $-14$ at $z=$1.3--2.6~\citep{alavi2016}. These luminosities are expected to trace galaxies in halos of masses $\sim1.5\mbox{--}4\times10^9\msun$\citep{fink15b,finl17}.The clustering of mass-selected galaxy samples indicates that star formation remains efficient down to halo masses of at least $10^{10.5}\msun$ at $z<12$~\citep{paquereau2025}, but this remains over $100\times$ more massive than the HI cooling limit. Despite its unprecedented sensitivity, JWST is unlikely to probe significantly more deeply than it has done, hence direct constraints on star formation in less-massive halos will remain out of reach apart from isolated strongly lensed systems~\citep[for example,][]{rb2022,morishita2025}.

Metal absorbers offer a complementary probe of low-mass galaxies because they are easier to detect in absorption than in emission~\citep{finl13,finl20,doughty2023}. A simple way to ask how many stars form in faint systems is to compare the total metal mass density in absorbers to the stellar mass density. If the metal density exceeds what is expected given the stellar mass density and a reasonable mean metal yield\footnote{defined as the IMF-averaged mass of metals ejected per mass divided by stellar mass in stars that do not explode in Type II SNe.}, then a contribution from fainter or obscured galaxies is implied~\citep{maddick14,perouxhowk2020}. Observations indicate that more than half of all metals at $z>3.5$ reside in absorption-selected systems~\citep{perouxhowk2020}, and for these the host galaxies are usually undetected. While this confirms that early star formation proceeded in faint systems, it offers little further insight into their nature.

The assumption that each absorber traces one galaxy opens up the possibility of abundance-matching. This involves identifying a geometric cross section that reconciles the galaxy number density with the absorber line incidence: If the comoving number density of host galaxies per stellar mass $M_*$ (that is, the stellar mass function SMF) is $dn/dM_*(M_*)$ and the proper geometric cross section for a galaxy's environment to be observed with column density greater than $N$ is $\sigma(M_*,>N)$, then the line incidence, defined as the cumulative mean number of absorbers observed with column density greater than $N$ per unit absorption path length $dX$, is~\citep[for example,][]{bahc69,sargent1979,young1982,bb1991}:
\begin{equation}\label{eqn:dndX}
\ell(>N) \equiv \frac{dn}{dX}(>N) = \frac{c}{H_0}\int_{M*,1}^\infty dM_* \frac{dn}{dM_*}\sigma(M_*,>N)
\end{equation}
where $M_{*,1}$ is the minimum stellar mass of galaxies that host absorbers. Equation~\ref{eqn:dndX} may be used in different ways. If $M_{*,1}$ (or, equivalently, the galaxy number density) is specified \emph{a priori}, then a measurement of $\ell(>N)$ constrains $\sigma(M_*,>N)$~\citep{sargent1979,young1982}. This approach is only as realistic as the adopted model for $\sigma(M_*,>N)$. To this end, many works assume that the absorbing region's linear size $R$ varies as a known power law of the host galaxy luminosity; that is, $R \propto L^\beta$~\citep{sargent1979}. Alternatively, if $\sigma(M_*,>N)$ is specified, then the total galaxy number density is constrained. If the latter is related to a (possibly extrapolated) SMF,  then $M_{*,1}$ is constrained, with implications for the abundance of low-mass galaxies. This degeneracy between host number density and absorption cross section is the key obstacle to using absorbers to constrain star formation in low-mass halos. It may be broken through direct observations of absorbers' host galaxies~\citep{bb1991}, but of course this is only possible for bright galaxies whose host halos' stellar populations are already constrained in other ways. 

In order to motivate this work, we argue that the simplest approach implied by Equation~\ref{eqn:dndX} is crude, unjustified, and potentially unphysical.
\begin{itemize}
\item It is crude because any statement of the line incidence $\ell$ suppresses heterogeneity in absorber column density. Such binning is necessary when only a handful of absorbers are available, but modern instrumentation now uncovers catalogs of $10^3$--$10^5$ systems or more~\citep{cook13,lan2020,hasa20,anand2025}. The statistical power contained in such catalogs calls for an analogue to Equation~\ref{eqn:dndX} that accounts naturally for absorber heterogeneity. Relatedly, observations indicate that a single galaxy's circumgalactic medium typically presents absorption spanning a range of strengths~\citep{stei10}. This clue is ignored when the cross section $\sigma(M_*,>N)$ consolidates heterogeneous absorbers into a single bin.
\item It is unjustified because the relationship between the geometric cross section to absorption and the stellar mass remain uncertain. As observations suggest that geometric cross section increases with luminosity~\citep{chur13}, this gap is often addressed by explicitly allowing the cross section to grow with luminosity in a prescribed way~\citep{holmberg1975,sargent1979,hasan2022}. However, such relationships have not been shown to emerge naturally from theoretical models.
\item It is potentially unphysical because it formally allows geometric cross sections that imply an impossible total ion mass density. This may be accounted for by requiring that the total ion mass density $\sigma N_i m_i \leq \frac{\Omega_b}{\Omega_M} y M_h$. Here, $\Omega_b$ and $\Omega_M$ are the cosmological baryon and matter density parameters; $\sigma$ is the halo's geometric absorption cross section; $N_i$ and $m_i$ are the column density and ion mass; and $y$ is the metal yield expressed as the ratio of metal mass formed per unit of baryon mass.\footnote{This maximal yield may also be viewed as the product of the star formation and metal production efficiencies.}
\end{itemize} 

These limitations, combined with the dramatic increase in data quality over the past fifty years, argue for revising how absorbers are matched to galaxies. We take a step in this direction by modeling the absorber column density distribution (CDD), which is simply the \emph{derivative} of Equation~\ref{eqn:dndX}, 

\begin{equation}\label{eqn:dndXdN}
\frac{d^2n}{dX dN} \equiv - d\mathscr{l}/dN = -\frac{c}{H_0}\int dM_* \frac{dn}{dM_*}\frac{d\sigma(M_*,>N)}{dN}
\end{equation}

The negative sign arises because the cumulative absorption cross section ($\sigma(M_*,>N)$, or CACS) generically decreases with $N$ but the CDD is conventionally represented as a histogram. Whereas Equation~\ref{eqn:dndX} depends on the CACS, Equation~\ref{eqn:dndXdN} depends on its derivative. This modification allows galaxies at fixed mass to host absorbers spanning a range of strengths, naturally leveraging the rich information contained in modern absorber catalogs. The additional realism necessitates a generalized model for $\sigma(M_*,>N)$ that characterizes how it varies independently with $N$ and $M_*$. Our goal is to extract such a model from realistic cosmological simulations but then apply it to interpret observations in such a way that results are not heavily dependent on the simulations. 



We begin in Section~\ref{sec:sims} by summarizing our cosmological simulations and our method for extracting simulated absorbers. We use a comparison between the predicted and observed SiIV CDD to support the simulation's overall realism and demonstrate convergence with respect to spectral resolution. We confirm that the simulation reproduces the observed galaxy stellar mass function at $z=5$--6. We expose the predicted stellar mass-halo mass relationship and test it against observational inferences and a complementary model at $z=6$. In Section~\ref{sec:method}, we review the virial incidence, or the line incidence of dark matter halos assuming that each halo presents a geometric cross section equal to its virial cross section. We show how it relates to the dark matter halo mass function (HMF) and confirm that the simulated HMF reproduces analytic expectations over the resolved range. As a first application, we use a comparison between the virial and absorber incidences to suggest that simulated CIV absorbers trace virialized gas while simulated SiIV, MgII, and OI absorbers contain an extravirial contribution. In Sec.~\ref{sec:res_sim}, we derive a general functional form for the CACS from our simulations. As a sanity-check, we then verify that characterizing individual galaxies' absorbing halos in this way accounts for the full CDD that emerges independently from ray-casting while confirming the presence of extravirial SiIV in the simulations. In Section~\ref{sec:empAbsGal}, we set the simulation aside and use our CACS model to study the observed abundance of SiIV absorbers at $z>5$. We first use the observed SiIV line incidence to argue that observed high-redshift SiIV systems trace predominantly virialized gas. We next perform a full fit to the CACS parameters in order to reconcile observations of the galaxy SMF with the SiIV CDD at $z>5$. We use the resulting fit parameters to quantify the likely relative contribution by galaxies of varying masses to the observed CDD and show that Holmberg-like power law relations between galaxy mass and absorption cross section appear disfavored. We show that the fraction of absorbers without visible counterparts will offer complementary constraints on the model. Finally, we discuss our results in Section~\ref{sec:discuss} and summarize in Section~\ref{sec:sum}. 

Throughout this work, column densities are denoted with $N$ in units of ions cm$^{-2}$, and logarithms are base-10. Our default cosmology is one in which ($\Omega_M, \Omega_\Lambda, \Omega_b, H_0, X_H$) = (0.3089, 0.6911, 0.0486, 67.74, 0.751), and our default stellar initial mass function is~\citet{krou01}.
 
\section{Simulation and Analysis}\label{sec:sims}
This section presents salient details of our simulation and describes how synthetic absorbers and galaxies are extracted from it.

\subsection{Simulation Details}\label{ssec:simDetails}
Our cosmological radiation hydrodynamic simulation uses an updated version of the Technicolor Dawn code baseline~\citep{finl18} to model a $(16.5\hmpc)^3$ volume using $2\times704^3$ gas and dark matter particles. For reference, this means that the dark matter halo mass function is resolved down to a 20-particle mass of $3.26\times10^7\msun$ and galaxies are resolved down to a 64-star particle mass of $8.2\times10^6\msun$. The radiation field is discretized spatially into $88^3$ uniformly sized voxels and spectrally into 32 evenly sized energy groups evenly sampling the range 1--10 Ryd.

With respect to our previous work~\citep{finl18}, our simulation's radiative transport 
solver incorporates upgrades to the stellar emissivity and the redshift-dependent ionizing 
escape fraction. We additionally incorporate treatments for re-processing of the metagalactic ionizing 
background by HeII ions. Full details are given in Section 2 
of~\citet{huscher24} but are not central to the current work.

\subsection{Synthetic Absorbers}\label{ssec:synthAbs}
We now describe our method for extracting a realistic catalog of metal absorbers from the simulation. We will only use this catalog to perform an internal consistency test of our model for the CACS in Section~\ref{sec:res_sim}; that is, it will not contribute to our main results in Section~\ref{sec:empAbsGal}.

\begin{figure}
\centering
\includegraphics[width=90mm]{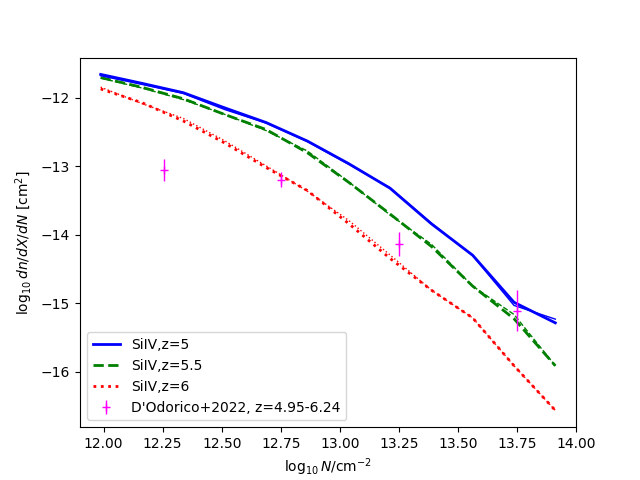}
\caption{Simulated SiIV CDDs at $z=$5--6 compared with observations by~\citet{dodo22} at $\langle z \rangle = 5.595$. As shown in that work, the SiIV CDD is well-resolved and in reasonable agreement with observations for column densities $\log(N/\mathrm{cm}^2) > 12.5$, where observational incompleteness is not an issue. Light and heavy curves illustrate the CDDs using the ``central $\sigma$" and ``mean $\sigma$" methods, respectively (see text). They are nearly coincident, confirming that SiIV predictions are converged with respect to spectral resolution. 
}
\label{fig:cddfs}
\end{figure}
Our method for extracting a catalog of synthetic metal absorbers follows our previous work~\citep{finl20} with a slight modification to address spectral resolution limitations. As before, we cast a sightline through the simulation snapshot, wrapping periodically at simulation boundaries. We divide the sightline into pixels spanning a velocity width 2 km s$^{-1}$ and compute the local density, metallicity, temperature, and proper motion at each pixel. We then compute the local ionization state under the assumption of ionization equilibrium adopting the local radiation field, temperature, and metallicity. We compute optical depth as a function of Hubble velocity using Voigt profiles~\citep{huml79}. Finally, we smooth the resulting intrinsic transmission curve with a Gaussian response function with a full width at half-maximum of 10 km s$^{-1}$ and add Gaussian random noise corresponding to a signal to noise of 50 per pixel. Absorbers are identified as regions where the mean flux, averaged over three consecutive pixels, drops more than 5$\sigma$ below unity. Absorbers identified within 50 km/s of one another are merged into ``systems"~\citep[for example,][]{dodo13}.

The difference with respect to previous work involves deriving the photoionization rate $\Gamma$ for a given species owing to the local radiation field. Absent spectral resolution limitations, this is the usual expression:
\begin{equation}\label{eqn:GammaReal}
\Gamma \equiv 4 \pi \int \sigma_\nu \frac{J_\nu}{h\nu} d\nu
\end{equation}
Here, $\sigma_\nu$ is the photoionization cross section at frequency $\nu$ and $J_\nu$ is the local specific intensity in ergs s$^{-1}$ cm$^{-2}$ Hz$^{-1}$ Sr$^{-1}$. If $J_\nu$ is modeled with \emph{finite} energy resolution, then the contribution $\Gamma_i$ to the total photoionization rate owing to the $i$th energy bin, centered at $\nu_i$ and spanning frequencies $\nu_{i-}$--$\nu_{i+}$, is
\begin{equation}\label{eqn:GammaFSR}
\Gamma_i \equiv \frac{4 \pi J_{\nu_i}}{h\nu_i} (\nu_{i+} - \nu_{i-}) \sigma_i 
\end{equation}
Here, $\sigma_i$ is the photoionization cross section that is applied to \emph{all} of the flux in the $i$th energy bin. Previously, we assumed $\sigma_i = \sigma_\nu(\nu_i)$; that is, that the photoionization cross section at the bin's central energy. This is reasonable as long as $\sigma_\nu$ is constant or varies linearly with energy across an energy bin. If it does not (for example, because the energy bin straddles a key photoionization threshold), then the method for choosing $\sigma_i$ becomes significant. As a test of our spectral resolution convergence, we have implemented the ability to set $\sigma_i$ equal to the mean of $\sigma_\nu$ over the energy bin:
\begin{equation}\label{eqn:sigmaMean}
\overline{\sigma_i} \equiv \frac{\int_{\nu_{i-}}^{\nu_{i+}} \sigma_\nu d\nu}{(\nu_{i+}-\nu_{i-})}
\end{equation}
By extracting our simulated absorber catalogs using both the previous ``central" method and the alternative ``mean" method, we estimate the impact of spectral resolution limitations. Figure~\ref{fig:cddfs} uses the resulting SiIV CDDs to show that, for our current energy bin width of 0.28125 Ryd, using $\overline{\sigma_i}$ rather than $\sigma_\nu(\nu_i)$ has no impact on SiIV.\footnote{The impacts on OI and MgII are likewise negligible at our spectral resolution, but the impact on CIV is substantial.} 
As the ``mean" method accounts qualitatively for cases in which an energy bin straddles an ionization threshold, we adopt it here even though Fig.~\ref{fig:cddfs} indicates SiIV predictions are insensitive to the choice.

\subsection{The Simulated Galaxy Stellar Mass Function}

\begin{figure}
\centering
\includegraphics[width=90mm]{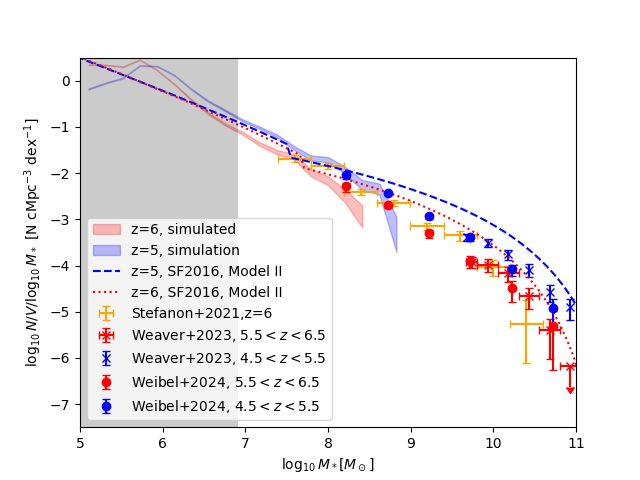}
\caption{Simulated SMFs at $z=5$ and $z=6$ (shaded regions), compared with observational inferences~\citep{weav22,weib2024,stefanon21}. We also show the predicted SMFs from combining the~\citet{sunf16} stellar baryon fractions with the Sheth-Tormen~\citep{shethTormen2001} halo mass function given our adopted cosmology (dotted curves). The shaded region on the left indicates the stellar mass range that our simulation does not resolve with at least 64 star particles. Agreement between simulations and observations is satisfactory around $10^{7.5}$--$10^{8.5}\msun$.
}
\label{fig:smf}
\end{figure}

Our method of using simulations to relate geometric absorption cross sections to galaxy stellar mass rests on the assumption that the stellar masses themselves are reasonable. We verify this in Figure~\ref{fig:smf}, which compares the simulated SMFs at $z=6$ and $z=5$ with recent observations~\citep{weav22,weib2024,stefanon21}. All observations are adjusted to our choice of cosmology and initial mass function as described in Appendix A. Simulated galaxies are extracted using {\sc skid}~\citep{governato1997,stadel2001}. Our simulations are in reasonable agreement with observed SMFs despite the weak overlap in dynamic range (the lack of massive galaxies simply reflects our limited simulation volume). As long as the adopted metal yields are realistic, the overall mass of metals produced is too. Figure~\ref{fig:cddfs} suggests that the overall SiIV mass is indeed plausible (for an observational test of the simulated galaxy metallicities, cf.~\citealt{Kusmic_2026}). The \citet{sunf16} stellar baryon fractions appear to overproduce the observed SMF even though the adopted cosmology and initial mass function are corrected to ours. The discrepancy probably reflects the fact that~\citet{sunf16} calibrated their model against observations of the rest-frame ultraviolet luminosity function, including an uncertain dust correction. To the extent that this overproduction is real, the~\citet{sunf16} baryon fractions underpredict galaxies' host halo masses.

\subsection{The Simulated Galaxy-Halo Relationship}\label{ssec:fstar}

\begin{figure}
\centering
\includegraphics[width=90mm]{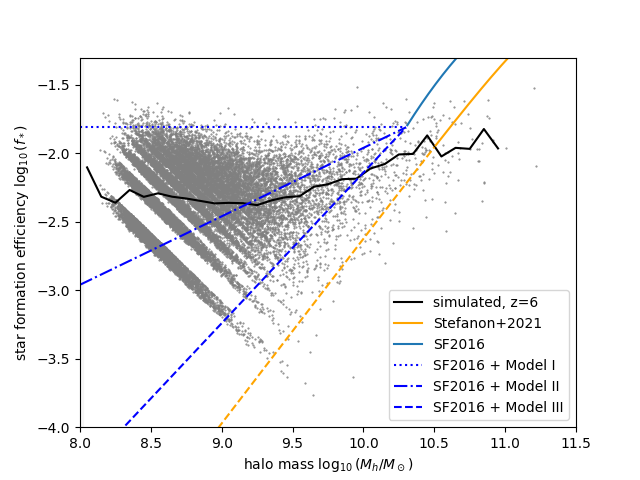}
\caption{The simulated stellar baryon fraction at $z=6$ as a function of host halo mass. Points represent individual simulated central galaxies; the black curve indicates their running median. Blue curves are from~\citet{sunf16}, converted to our cosmology and IMF (see the Appendix). The orange curve is from~\citet{stefanon21}, adjusted to our choice of a Kroupa IMF. Our simulated trend roughly resembles the Model II extrapolation of~\citet{sunf16}. The grey diagonal stripes are a resolution artifact indicating simulated halos that contain small numbers of star particles.
}
\label{fig:fstar_mhalo}
\end{figure}

We next map from stellar mass to halo mass. We compute each simulated galaxy's host halo mass by growing a sphere about its center of mass until the enclosed density drops below the redshift-dependent virial overdensity~\citep{brya98}. The resulting stellar mass-halo mass relationship at $z=6$ is shown in Figure~\ref{fig:fstar_mhalo}, where the y-axis quantifies the stellar baryon fraction $f_{*,b} \equiv \frac{M_*}{M_h}\frac{\Omega_M}{\Omega_b}$. The median simulated trend is flat at low masses and then grows at most shallowly for $\log(M_h/\msun) > 9.5$. Comparison with empirical modeling by~\citet{sunf16} (blue curves) reveals that the median simulated trend is roughly consistent with their Model II extrapolation to low masses, with the full range of star formation efficiencies bounded by their Models I and III\footnote{The fit parameters given in the original publication are replaced with those in Appendix B.}. The simulated efficiency is $\approx 3 \times$ higher than an extrapolation of the~\citet{stefanon21} efficiency, but this is not small compared to systematic uncertainties (see, for example, their Figure 13). For halos whose central galaxies are observable ($\log(M_h/\msun)>10.5$), the~\citet{stefanon21} efficiency falls systematically below the~\citet{sunf16} inference, consistent with the suggestion in Figure~\ref{fig:smf} that the~\citet{sunf16} efficiencies overproduce the observed SMF.


\section{Matching Galaxies and Absorbers to Halos Using the Virial Line Incidence}\label{sec:method}

In order to derive suitable values for the fit parameters from simulations, we must identify the radius out to which absorbing gas is associated with the central galaxy. Defining such a boundary is necessary because Equations~\ref{eqn:dndX}--\ref{eqn:dndXdN} do not refer explicitly to the gas's radial profile. We will eventually test this choice by comparing results with ray-casting. We begin by asking whether simulated absorbers could arise primarily within halos. A simple way to do this is to compare the simulated $\ell$ that results directly from ray-casting against the simulated ``virial incidence" $\ell_v$ where the latter is the line incidence if each simulated dark matter halo absorbs with unity covering fraction out to its virial radius (see also Section 2 of~\citealt{hasan2022}). Absent dynamic range limitations, the full virial line incidence is
\begin{equation}\label{eqn:dndXVirTheory}
\ell_{v,\mathrm{full}} \equiv \frac{c}{H_0} \int_0^\infty dM \frac{dn}{dM}(M,z) \pi r_{V}^2(M,z)
\end{equation}
where $dn/dM(M,z)$ is the dark matter halo mass function and $r_V(M,z)$ is the proper virial radius of a halo of mass $M$ at redshift $z$.

When interpreting simulations, this integral reduces to a sum over simulated galaxies or halos:
\begin{equation}\label{eqn:dndXVirSim}
\ell_{v,\mathrm{sim}} \equiv \frac{c}{H_0 V}\sum_i \pi r_{V,i}^2
\end{equation}
Here, $i$ runs over simulated systems; $V$ is the comoving simulation volume; and $r_{V,i}$ is the radius of the smallest region about each simulated galaxy where the overdensity falls below the virial overdensity~\citep{brya98}.

\begin{figure}
\centering
\includegraphics[width=90mm]{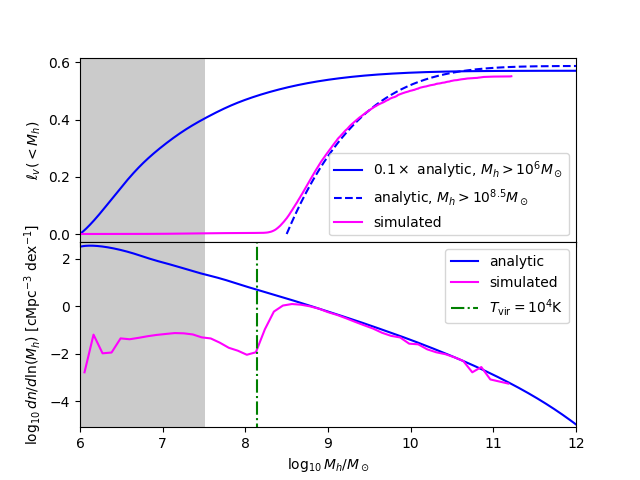}
\caption{(Top) The cumulative virial line incidence $\ell_v$, defined so that the $y$-axis gives the line incidence~\ref{eqn:dndX} of dark matter halos less massive than the value on the $x$-axis. Solid and dashed blue curves indicate the analytic prediction using the Sheth-Tormen mass function for halos more massive than $10^6$ and $10^{8.5}\msun$ (Equation~\ref{eqn:dndXVirTheory}), respectively, while the solid magenta curve indicates the virial line incidence including all simulated halos that host galaxies (Equation~\ref{eqn:dndXVirSim}). (Bottom) The Sheth-Tormen mass function and the simulated mass function of galaxy-hosting halos. Both panels refer to $z=6$. Shaded regions indicate the unresolved mass range.
}
\label{fig:ell}
\end{figure}

For reference, we compare the theoretical and simulated $\ell_v$ and dark matter halo mass functions at $z=6$ in Figure~\ref{fig:ell}. The theoretical mass function and line incidences are computed from the Sheth-Tormen mass function. Simulated curves account only for dark matter halos that host galaxies. Comparing the solid and dashed blue curves in the top panel indicates that the theoretical $\ell_v$ depends sensitively on the range of halo masses considered. The solid curve accounts for all halos more massive than $10^6\msun$ and is scaled down by $10\times$ to fit in the figure. A typical sightline traverses many low-mass halos as they are abundant, leading to a large $\ell_v$. They will contribute few metal absorbers, however, because star formation is inefficient in halos less massive than $10^8$--$10^9\msun$~\citep{thoul96,okam08,finl11b,nebrin23,dhandha2025}. To account for this, the dashed blue curve shows the theoretical $\ell_v$ when considering only halos more massive than $10^{8.5}\msun$; this $\ell_v$ is roughly $10\times$ lower. The good agreement with the simulated (magenta) curve confirms that our simulation captures most of the relevant halos.

The bottom panel shows that the the simulated halo catalog is in good agreement with the Sheth-Tormen mass function for masses between $10^{8.2}\msun$--$10^{11}\msun$. Less-massive halos are missing from the simulation because they cannot grow galaxies even if they are resolved numerically. More-massive halos are missing because they are too rare for our limited cosmological volume~\citep{bark04}. In between these limits, the excellent agreement supports our spherical-overdensity method for associating galaxies to host halos as well as the stellar mass-halo mass relation in Figure~\ref{fig:fstar_mhalo}.

\begin{figure}
\centering
\includegraphics[width=90mm]{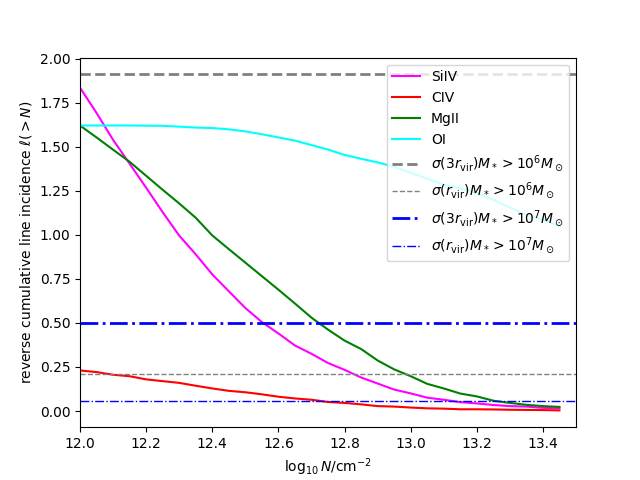}
\caption{The reverse cumulative line incidence of simulated metal absorbers at $z=6$ as a function of minimum column density for several different ions as compared to the line incidence that results from assuming that all simulated galaxies absorb with unity covering fraction out to 1 and 3.0 virial radii for two different cuts in minimum stellar mass. Comparing the solid curves with the blue dot-dashed and grey dashed curves reveals that, in our cosmological simulation, weak SiIV and MgII absorbers and essentially OI absorbers trace predominantly extravirial gas.
}
\label{fig:dndX_logN}
\end{figure}

Having confirmed that our simulated $\ell_v$ matches theoretical expectations for atomically cooled halos, we now show that simulated SiIV absorbers likely trace a mixture of virialized and extravirial gas, with weak systems receiving a larger extravirial contribution. Figure~\ref{fig:dndX_logN} compares the predicted SiIV absorber line incidences to the virial incidence that arises when considering only the halos of simulated central galaxies at $z=6$. Simulated virial incidences from host halos of galaxies with stellar mass greater than $10^6$ and $10^7\msun$ are indicated with thin blue short-dashed and gray dashed curves, respectively. Corresponding thick curves indicate the result when $\ell_v$ is re-computed using three times the virial radius. Broadly, strong SiIV absorbers may trace halo gas (as previously noted for CIV by~\citealt{hasan2022}) while weak absorbers must trace extravirial gas. As before, inferences depend on the choice of minimum stellar mass. For example, virialized gas associated with central galaxies more massive than $M_*=10^6\msun$ could account entirely for SiIV absorbers stronger than $10^{12.8}$cm$^{-2}$. If we attempt instead to attribute simulated absorbers to galaxies with stellar mass $>10^7\msun$, then halo gas can only account for SiIV absorbers stronger than $10^{13.0}$cm$^{-2}$. In this case, weaker absorbers require a contribution from extravirial gas.

The inference from Figure~\ref{fig:dndX_logN} that simulated metal ions partially trace extravirial gas depends only on $\ell$. Applying our reasoning to other ions reveals that simulated SiIV, MgII, and OI absorbers are much more likely to trace extravirial gas at $z=6$ than CIV (for observational support that high-redshift MgII absorbers trace extravirial gas, see~\citealt{bord24}). For the same reason, however, we do not use Figure~\ref{fig:dndX_logN} to determine whether \emph{observed} absorbers trace extravirial gas unless their $\ell$'s agree. For example, a simulated overproduction of weak SiIV absorbers (\citealt{dodo22} and Figure~\ref{fig:cddfs}) would exaggerate the case for extravirial SiIV. We will address the origin of observed SiIV absorbers directly using the observed $\ell$ in Figure~\ref{fig:dndX_logN_obs_z5}.

\section{The Simulated Absorber-Galaxy Relationship}\label{sec:res_sim}
We now analyze the simulated mean SiIV absorption profiles and relate these to ensemble absorber statistics. To derive mean absorption profiles, we collapse the local ion densities contained within a cubical volume of length $l_{\mathrm{cube}}=2r_{\mathrm{v}}$ centered about each simulated galaxy into a two-dimensional grid with 128 pixels per side. We then compute each galaxy's predicted CACS considering only pixels that lie within one virial radius of the galaxy. Physically, this means that we consider all gas originating in a cylinder with radius equal to the virial radius and length equal to twice the radius. By averaging results over many simulated galaxies within narrow mass bins, we then construct the average CACS over a range of stellar mass and column density. The results indicate how galaxy abundance and CACS are predicted to contribute to the ensemble statistics.

\begin{table}
\caption{SiIV CACS fit parameters for Equations~\ref{eqn:sigmaFit}}
\centering
\begin{tabular}{l | c c c c c c}
\hline
z & $s_0$ & $s_1$ & $l_0$ & $l_1$ & $p_0$ & $p_1$ \\
\hline
\hline
\multicolumn{7}{c}{simulated} \\
6 & 0.718 & 0.032 & 8.841 & 0.563 & 70.356 & 0.069 \\
\hline
\multicolumn{7}{c}{empirical} \\
5.595 &$-2.6^{+5.5}_{-4.5}$ & $1.8^{+0.9}_{-0.9}$ & $8.9^{+0.54}_{-0.35}$ & $0.51^{+0.066}_{-0.062}$ & $35^{+200}_{-170}$ & $21^{+27}_{-24}$\\
\end{tabular}
\label{tab:xFits}
\end{table}

\subsection{Simulated SiIV Absorption Profiles}\label{ssec:sigmaMean}
\begin{figure}
\centering
\includegraphics[width=90mm]{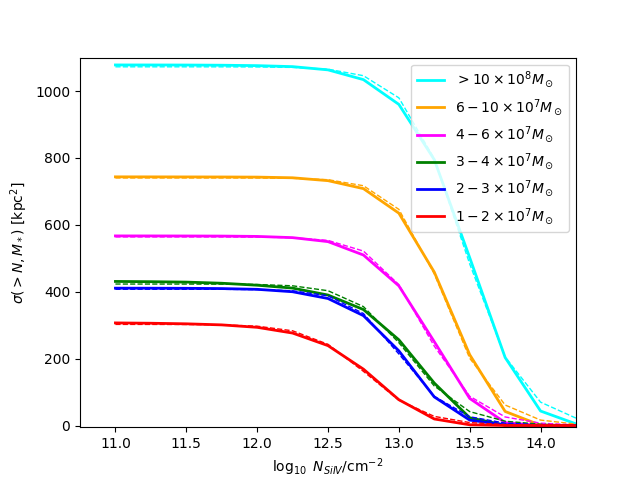}
\caption{The simulated SiIV CACS in six bins of stellar mass at $z=6$. Only gas within the virial radius is considered. Heavy solid and light dashed curves indicate the simulated trend and the fit from Equation~\ref{eqn:sigmaFit}.
}
\label{fig:sigmaN}
\end{figure}

Figure~\ref{fig:sigmaN} illustrates how the SiIV CACS varies separately with column density and stellar mass at $z=6$. Intuitively, galaxies contribute very little absorption in regions where the CACS is flat (for example, at very low columns). By contrast, they contribute the most absorption in regions where the CACS declines steeply with $N$. Several trends are evident in Figure~\ref{fig:sigmaN}. First, the normalization increases with stellar mass such that more-massive galaxies are associated with overall larger CACS. This is qualitatively consistent with the observation that massive galaxies are more likely to host absorbers at fixed minimum strength~\citep{burchett2016}. Second, for a given stellar mass, there is a threshold column density below which the predicted covering factor within the virial radius is unity. This threshold column density increases with stellar mass. Third, the power law slope that describes how the CACS drops with increasing columns steepens with mass. 

These CACS profiles are reasonably well fit using a van Genuchten-Gupta fit function~\citep{vang93}  in which the fit parameters have a linear dependence on $\log(M_*/\msun)$:
\begin{align}
\sigma(M_*,>N) & = \frac{s(M_*) \pi r_{V}(M_*)^2}{1 + \left(\frac{\log(N)}{\log({N_0})(M_*)}\right)^{P(M_*)}} \label{eqn:sigmaFit} \\
s(M_*) & = s_0 + s_1 \log(M_*/\msun) \nonumber \\
\log(N_0)(M_*) & = l_0 + l_1 \log(M_*/\msun) \nonumber \\
P(M_*) & = p_0 + p_1 \log(M_*/\msun) \nonumber
\end{align}
Parameters $s_0$ and $s_1$ set the ratio of maximum cross section to the virial cross section and are dimensionless. $l_0$ and $l_1$ set the threshold column density below which the CACS saturates to this level and have units cm$^{-2}$. $p_0$ and $p_1$ set the slope of the high-column cutoff and are dimensionless. Stellar masses are in units of $M_\odot$. Procedurally, we first fit for $s$, $\log(N_0)$, and $P$ for different stellar mass ranges as shown in Figure~\ref{fig:sigmaN}. We then determine the stellar mass dependence of those fit parameters. For reference, fully self-similar profiles~\citep{chur13} correspond to $s_1=p_1=l_1=0$. 

Comparison between the dashed (fitted) and solid (simulated) curves in Figure~\ref{fig:sigmaN} confirms that our proposed fit function accommodates the theoretical CACS profiles well. In detail, close inspection reveals that the fits systematically overproduce the simulated profiles at large column densities, particularly for massive galaxies. This discrepancy results from the unweighted likelihood function that we use in the fitting. If $\sigma_{s,i}$ and $\sigma_{m,i}$ represent the simulated and modeled cross-sections in column density bin $i$, then minimizing the likelihood $\sum_i(\sigma_{s,i}-\sigma{m,i})^2$ naturally permits larger \emph{fractional} errors where the cross sections are smaller. A suitable weighting scheme could improve the fits at large column densities, but only by degrading the fits at lower column densities. For the purpose of demonstrating our method, we therefore we retain the unweighted likelihood. The associated simulated SiIV CACS fit parameters are given in Table~\ref{tab:xFits}.
\begin{figure}
\centering
\includegraphics[width=90mm]{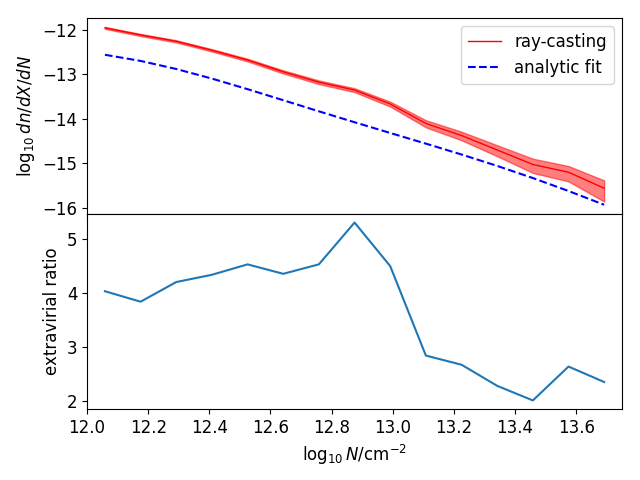}
\caption{The simulated SiIV CDD at $z=6$ derived in two independent ways. The red shaded region indicates predictions from a ray-casting approach that simulates observations directly; its thickness gives the $\sqrt{N}$ confidence interval. The blue dashed curve indicates the result from combining Equations~\ref{eqn:sigmaFit} and~\ref{eqn:dndXdN}. Agreement is nontrivial because the methods used to derive the CDDs are completely independent.
}
\label{fig:dndXdN}
\end{figure}

As an internal consistency check, we now test whether our method reproduces the full CDD that results directly from ray-casting. We do this by inserting the fits from Equation~\ref{eqn:sigmaFit} into Equation~\ref{eqn:dndXdN}, using the parameters in Table~\ref{tab:xFits}. The integral is evaluated as a sum over all simulated galaxies including unresolved systems. Host halo masses are inferred from simulated stellar masses using~\citet{sunf16}. Here and throughout the rest of our study, we adopt their Model II extrapolation to low masses as we have shown that it agrees reasonably well with our simulations (Fig.~\ref{fig:fstar_mhalo}). This test is nontrivial: First, the parameters in Equation~\ref{eqn:sigmaFit} are not adjusted to match the simulated CDD; instead, they are calibrated to match the simulated geometric cross-section profiles. Second, our method for deriving CACS profiles involves stacking the simulated three-dimensional column density profiles in one direction and considering only pixels that lie out to the host halo's virial radius, effectively including all gas lying in a cylinder centered on the galaxy's position. This could misrepresent absorbing gas geometries that are extended or asymmetric. Third, our method assumes that each galaxy hosts one isolated absorbing cloud; it does not account for the possibility that stronger absorbers may arise where metal clouds from neighboring galaxies or central-satellite pairs overlap.

We show in Figure~\ref{fig:dndXdN} (top panel) that the resulting SiIV CDD at $z=6$ is in marginal agreement with the one that emerges directly from ray-casting at the strong end while underproducing it at the weak end. The discrepancy between the two derived CDDs, indicated as a ratio in the bottom panel, quantifies the contribution from extravirial gas in a way that complements Figure~\ref{fig:dndX_logN}. Both figures indicate that up to 80\% of the simulated weak absorbers and 50\% of strong ones trace gas that lies outside of halos. Neither figure quantifies the relative extravirial contribution as a function of halo mass, however: the extravirial absorbers could equally originate in extensive clouds about rare, massive systems or confined clouds about abundant, low-mass ones. Apart from this discrepancy, however, the reasonable internal consistency between these two independent methods for computing the simulated CDD supports both methods.

\subsection{The Simulated SiIV-Galaxy Relationship}\label{ssec:simSiIVgal}
\begin{figure}
\centering
\includegraphics[width=90mm]{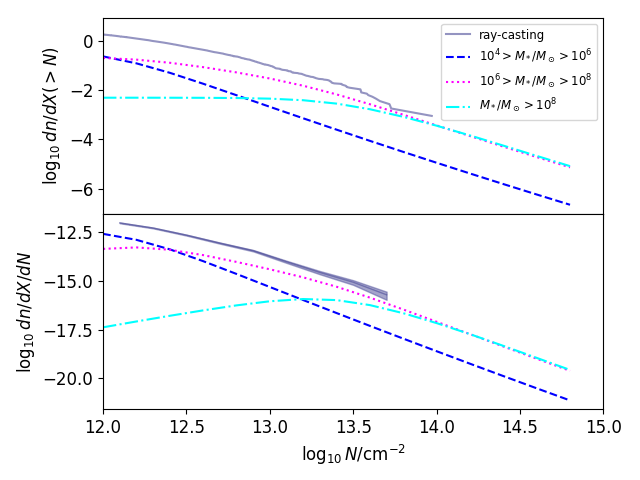}
\caption{Top: the simulated SiIV cumulative line incidence from ray-casting (grey shaded regions) as compared to the reconstruction using our analytic model at $z=6$. The simulated region includes all absorbing gas while the analytic reconstructions include only virialized gas for several different cuts in stellar mass. Bottom: the corresponding simulated CDD compared to the reconstructed CDDs adopting the same stellar mass cuts. Eliminating low-mass galaxies reduces the CDD at all masses, but the effect is larger at lower column densities.
}
\label{fig:dndXdN_Mlim}
\end{figure}
A central goal is to infer how galaxies of different masses contribute to absorbers of different column densities. The analytic fit in Equation~\ref{eqn:sigmaFit} enables us to address this by limiting the range of integration in Equation~\ref{eqn:dndXdN}. We illustrate this in Figure~\ref{fig:dndXdN_Mlim}. We generate this figure by evaluating Equations~\ref{eqn:dndX} and~\ref{eqn:dndXdN} as a sum over simulated galaxies when restricted to different stellar mass ranges as indicated in the legends. We use the fitted mean cross section from Equation~\ref{eqn:sigmaFit} with the fit parameters in Table~\ref{tab:xFits} for each simulated galaxy.

The top panel shows that, as one moves from weak absorbers to strong ones, the cumulative absorber line incidence becomes dominated by galaxies that are systematically more massive even though galaxies of all masses host absorbers of all column densities. Comparing the ray-casting line incidence (long-dashed gray curve) to the analytic prediction reveals that the analytic model underpredicts simulated SiIV absorbers, particularly at low column densities. This was seen previously in Figure~\ref{fig:dndXdN}, but the decomposition into different mass bins shows how extravirial metals about either low-mass or massive halos could supply the bulk of the weak absorbers.

The trends become even clearer in the bottom panel, where we show the CDD derived using Equation~\ref{eqn:dndXdN} using galaxies in different mass ranges. As before, galaxies less massive than $10^7\msun$ host relatively few strong absorbers while galaxies more massive than $10^8\msun$ contribute predominantly strong absorbers. This qualitative tendency for simulated low-mass galaxies to present negligible cross section for strong absorption has previously been noted by~\citet{keat16}, who reported that increasing a simulation's mass resolution at fixed cosmological volume boosts the predicted abundance of weak systems without affecting stronger ones. 

In summary, we have shown that Equation~\ref{eqn:sigmaFit} accommodates well the simulated CACS profiles; that incorporating the resulting fit parameters yields reasonable agreement with the CDD and $\ell$ that result directly from ray-casting; and that a weak association between strong absorbers and massive galaxies can be derived from the resulting fit functions even though galaxies at fixed mass host absorbers across a wide range of column densities. We have also argued that some simulated SiIV absorbers trace extravirial gas, particularly at the weak end of the CDD. Having thus motivated and explored our analytic model using simulations, we now set the simulation aside and apply the framework that we have derived from it directly to observations.

\section{The Empirical Absorber-Galaxy Relationship}\label{sec:empAbsGal}

We now apply the analytic CACS model that we introduced in the foregoing discussions to interpret the observed absorber-galaxy connection. We begin in Section~\ref{ssec:empAbsGal-priors} by revisiting the stellar mass-halo mass relation and the possible contribution of extravirial gas. We then introduce our priors. Having set up the model, we then present constraints on the SiIV CACS in Sections~\ref{ssec:empAbsGal-CACS}--\ref{ssec:empAbsGal-NMax}.

\subsection{Empirical Analysis: Method and Priors}\label{ssec:empAbsGal-priors}
\begin{figure}
\centering
\includegraphics[width=90mm]{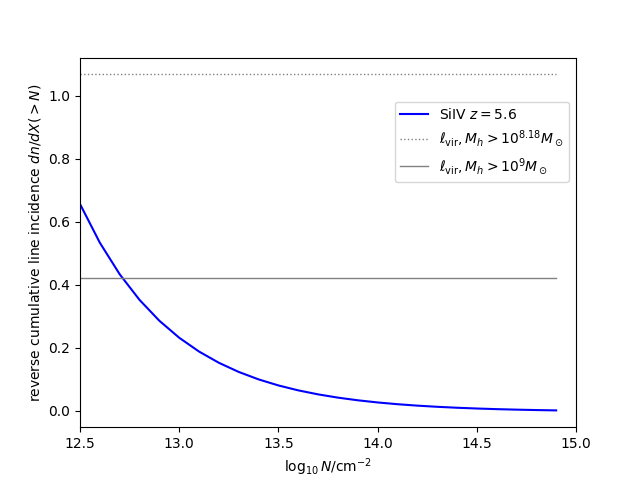}
\caption{The reverse cumulative line incidence of observed SiIV absorbers at $z=5.6$~\citep{dodo22}, compared to $\ell_\mathrm{vir}$ for two different minimum dark matter halo masses. Observed SiIV absorbers may arise entirely from virialized gas within atomically cooled halos ($\log(M_h/\msun)>8.18$; dotted curve).
}
\label{fig:dndX_logN_obs_z5}
\end{figure}

Our simulated fits implicitly assume the simulated stellar mass-halo mass relation. This could be incorrect, and in any case it is limited to the range of halo masses that our simulation captures. We address this by combining the theoretical (Sheth-Tormen) dark matter halo mass function with the observed stellar mass-halo mass relation~\citep{sunf16} including their Model II extrapolation to low stellar masses.

The possible need for extravirial gas arises because, while our simulated fits considered only virialized gas, Figure~\ref{fig:dndX_logN} suggested that some simulated SiIV absorbers are extravirial. In order to test whether extravirial SiIV is also implied \emph{observationally}, we compare in Figure~\ref{fig:dndX_logN_obs_z5} the observed cumulative line incidence of SiIV~\citep{dodo22} at $\langle z\rangle=5.595$ with the virial incidences of dark halos more massive than $10^{8.18}$ and $10^9\msun$. The lower of these two indicates a virial temperature of $10^4$K, roughly the hydrogen-cooling threshold at this redshift~\citep{nebrin23,dhandha2025}. The SiIV line incidence results from fitting power laws to the redshift-dependent CDDs in Figure 3 of~\citet{dodo22} and then interpolating to $\langle z\rangle=5.595$. The virial incidences are computed using Equation~\ref{eqn:dndX} along with the analytic dark matter halo mass function. Comparison indicates that the line incidence of observable SiIV absorbers ($\log(N) \leq 12.5$) is consistent with emerging entirely from virialized gas within hydrogen-cooling halos. Complementary to low-redshift observations indicating that CIV is never detected beyond a virial radius~\citep{chen2001,bordoloi2014,liangchen2014}, this supports the plausibility of galaxy-absorber matching scenarios in which each absorber's metals emerge predominantly in one host halo. Hence while our model will support the possibility of extravirial SiIV, we expect it to be subdominant.

The general absorption profile in Equation~\ref{eqn:sigmaFit}, when applied to joint observations of the SMF and the CDD, enables empirical inferences regarding the underlying CACS parameters. As a proof of concept, we apply our method to the SiIV CDD observed at $\langle z \rangle=5.595$ by~\citet{dodo22}. We adopt flat, linear priors on all six CACS parameters (Equation~\ref{eqn:sigmaFit}) and vary them independently over a uniform grid of values, sampling the parameter space at 22 points in each dimension.  These priors are given in Table~\ref{tab:priors}. We reject as unphysical models that fail any of the following checks:
\begin{enumerate}
\item $s$ and $p$ must be positive everywhere although negative values for $s_0, s_1$, and $p_0$ are permitted. \label{req:sp}
\item $p<500$ (this prevents floating point overflow). \label{req:fpo}
\item the total ion yield for each modeled galaxy mass may not exceed $y_\mathrm{max}$. \label{req:ymax}
\end{enumerate}

The total ion yield is defined as the mass in SiIV divided by the stellar mass. We restrict this ratio to fall below $10^{-3.1}$, which corresponds to a SiIV/Si fraction of roughly 50\% (for example, Table 3 of~\citealt{finl18}). We use the standard $\chi^2$ definition $\chi^2 = \sum_i (d_i-m_i)^2/\sigma_i^2$ to compute the log-likelihood for each model, where $d_i$ and $m_i$ are the measured and modeled CDDs at column density $i$, $\sigma_i$ is the reported uncertainty, and the sum is over all four data points reported by~\citet{dodo22}. This analysis differs fundamentally from the one that was used to obtain the theoretical CACS parameters: whereas the simulated CACS is extracted directly from the known galaxy-CGM relationship without reference to the CDD, the empirical one is derived indirectly by identifying the range of CACS models that reproduce the observed CDD.

\subsection{Empirical Analysis: Results}\label{ssec:empAbsGal-CACS}

\begin{table}
\caption{Priors for CACS fit parameters}
\begin{tabular}{l|c|l} 
\hline
parameter & range & function\\
\hline
\hline
$s_0$ & -10--5 & cross section: normalization \\
$s_1$ & -2.5--4.0 & cross section: mass dependence \\
$l_0$ & 7--15 & threshold column density: normalization \\
$l_1$ & 0.1--1.25 & threshold column density: mass dependence\\
$p_0$ & -250 -- 350 & high column density cutoff: normalization\\
$p_1$ & -51 -- 81 & high column density cutoff: mass dependence\\
\end{tabular}
\label{tab:priors}
\end{table}

\begin{figure*} 
\centering
\includegraphics[width=180mm]{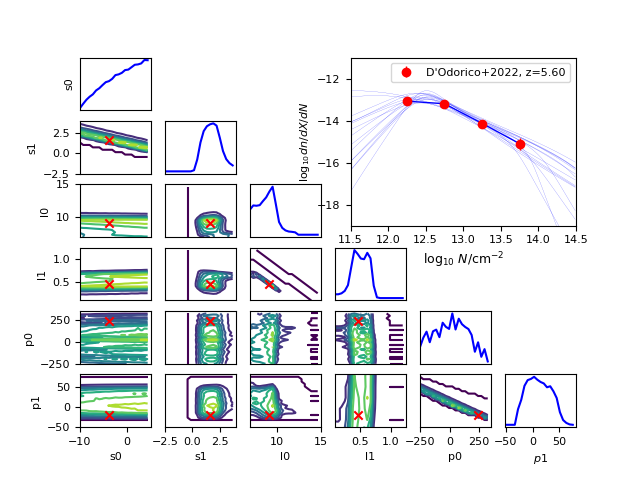}
\caption{Corner Plot: Constraints on the six parameters (Section~\ref{ssec:sigmaMean}) that describe the SiIV absorption profiles at $z=5.595$. Red \textcolor{red}{x}'s indicate the best-fit model. Top Right: The best-fit model (blue heavy curve) and a selection of models drawn randomly from the 67\% confidence region as compared to the~\citet{dodo22} CDD at $\langle z \rangle=5.595$. The model permits excellent fits, but extrapolations outside the constrained range diverge. The normalizations ($s_0$, $l_0$, and $p_0$) and slopes ($s_1$,$p_1$,and $l_1$) are anticorrelated.  Physical interpretations are given in the text.
}
\label{fig:dndXdN_emp_SiIV}
\end{figure*}

The heavy blue curve in the top-right panel of Figure~\ref{fig:dndXdN_emp_SiIV} confirms that the best-fit model yields excellent agreement with the observed CDD, with a total $\chi^2$ of 0.0181 when summed over the four data points. The light curves show a set of models randomly selected from the 1$\sigma$ confidence region in the full posterior. Unsurprisingly, the best-fit models match the measurements but diverge in regions where the CDD is unconstrained. At the weak end, some models predict a flattening while others predict a turnover. At the strong end, some models likewise predict a turnover whilst others continue to even larger columns at constant slope. The corner plot shows the covariance between different combinations of parameters, with red x's indicating the best-fit model in the top-right figure (heavy blue). The mean and 67\% confidence intervals derived from the marginalized one-dimensional posterior probabilities are given in Table~\ref{tab:xFits}. This analysis yields the following takeaways:
\begin{itemize}
\item Negative values of $s_0$ are permitted, consistent with the suggestion in Figure~\ref{fig:dndX_logN_obs_z5} that extravirial gas is not required to account for SiIV absorptions. 
\item $s_1$ is constrained to be positive, indicating galaxies that are more massive host larger absorbing clouds in units of $\pi r_V^2$.
\item $s_0$ and $s_1$ are anti-correlated: if the cross section normalization $s0$ is larger, then the cross section must increase less strongly with halo mass in order to avoid overproducing observed absorption systems.
\item The normalization of the characteristic column density $l_0$ falls well below the observed range, favoring models in which the covering fractions rise steeply to low column densities. 
\item The threshold column's mass dependence $l_1$ is weak but nonzero, consistent with the tendency for more massive galaxies to host predominantly stronger absorbers. 
\item $l_0$ and $l_1$ are anti-correlated: if the threshold column density $l_0$ is lower, then the mass-dependence $l_1$ must be larger.
\item The CACS slope at high columns $p = p_0 + p_1\log(M_*/\msun)$ is steep. A strong mass dependence is required so that massive systems generate more absorbers at all columns. For the best-fit model, $p$ varies from 200-500 for observable galaxies. 
\item $p_0$ and $p_1$ are degenerate such that a higher normalization requires a steeper mass dependence.
\item The fact that both $s_1$ and $l_1$ are constrained to be positive conflicts with observational evidence that absorption profiles are self-similar~\citep{chur13}.
\end{itemize}

\subsection{The Inferred Conditional Column Density Distribution}\label{ssec:ccdd}

\begin{figure}
\centering
\includegraphics[width=90mm]{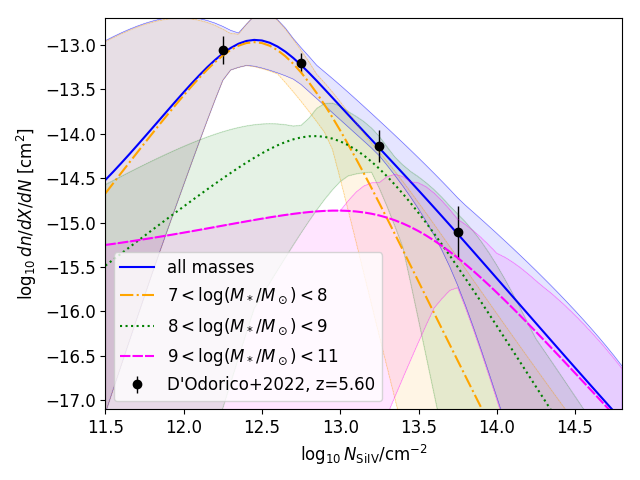}
\caption{The inferred conditional column density distribution for SiIV absorbers at $z=5.595$ based on the modeling in Figure~\ref{fig:dndXdN_emp_SiIV}. Thick curves and shaded regions represent the best-fit model and 67\% confidence range, respectively. Stronger absorbers generally trace massive galaxies while weak ones trace low-mass ones even though galaxies of all masses host absorbers of all strengths.
}
\label{fig:dndXdN_Mlim_emp_SiIV}
\end{figure}

Having derived a model for the SiIV CACS that reconciles the observed SMF, the underlying theoretical dark matter halo mass function, and the observed SiIV CDD, we ask how galaxies of different masses contribute to the observed CDD. Similarly to Figure~\ref{fig:dndXdN_Mlim}, we do this by restricting the range of integration in Equation~\ref{eqn:dndXdN} and deriving the resulting ``conditional column density distribution" (CCDD); that is, the CDD subject to the condition that the host galaxy's stellar mass falls within a restricted range. We show the result in Figure~\ref{fig:dndXdN_Mlim_emp_SiIV}. Here, the solid curve reproduces the best-fit total CDD from Figure~\ref{fig:dndXdN_emp_SiIV} while the other curves indicate the associated CCDDs. Shaded regions indicate the ranges spanned by representative models sampling the 67\% confidence region in the posterior, also shown in Figure~\ref{fig:dndXdN_emp_SiIV}.

Focussing just on the curves from the best-fit model, we see that stronger absorbers are generally associated with more massive galaxies. Our modeling does not, however, require this association to be tight: while observable galaxies ($\log(M_*/\msun) > 8$;~\citealt{stefanon21,weib2024}) predominantly host observable absorbers ($\log(N) > 13.2$;\citealt{davi23}), unobservably faint galaxies contribute significantly throughout the column density range where absorbers have been detected ($N <10^{14}$cm$^{-2}$). Conversely, observable galaxies can host absorbers with column densities falling throughout the observable range. The possibly dominant contribution of faint galaxies to strong absorbers is consistent with the recent non-detections of host galaxies about strong OI (\citealt{wu2023};$z>5$) and (at $z=$1--1.5) CIV and MgII~\citep{schroetter2021} absorbers, although~\citet{bord24} report an observational association between strong MgII absorbers and massive galaxies.

Switching our focus from the best-fit model to the range spanned by models that fall within the 67\% confidence region in the posterior, we see that the CCDD is not well-constrained by the CDD alone. This loose association between stellar mass and absorber strength indicates that the degeneracy between geometric cross section and host mass inherent in Equation~\ref{eqn:dndX}--\ref{eqn:dndXdN} remains within our model. At the same time, it emphasizes how absorption- and emission-selections may be complementary: our modeling strongly suggests that absorbers weaker than $10^{13}\mathrm{cm}^{-2}$ are dominated by unobservably faint galaxies ($M_*<10^8\msun$).

\begin{figure}
\centering
\includegraphics[width=90mm]{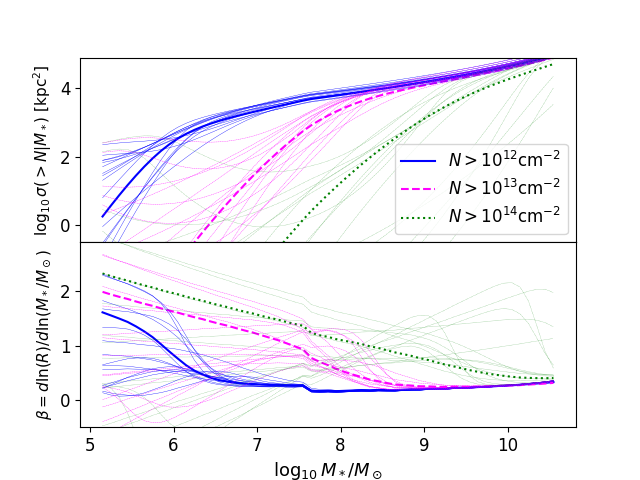}
\caption{\emph{(Top:)} The best-fit differential SiIV cross-section as a function of stellar mass and the logarithmic derivative of the absorption radius with respect to stellar mass\emph{(bottom)}. Heavy and light curves indicate the best-fit model and other models randomly chosen from the 67\% confidence region in the posterior (cf.\ Figure~\ref{fig:dndXdN_emp_SiIV}).
}
\label{fig:sigma_v_M_empirical_SiIV}
\end{figure}

The fitted differential cross sections in Figure~\ref{fig:dndXdN_emp_SiIV} enable us to ask whether observations support a simpler dependence of cross section on stellar mass than the one proposed in Equation~\ref{eqn:sigmaFit}. In particular, previous studies have invoked the \emph{Ansatz} that the geometric size of an absorbing region $R$ varies as a simple power-law function of host galaxy luminosity $L$; that is, $R \propto L^\beta$~\citep[for example,][]{hasan2022}. Our six-parameter CACS model differs conceptually from this two-parameter scenario in that it makes no assumption regarding the absorbing region's shape or relative location with respect to the host galaxy. However, if we assume circular symmetry with a unity covering factor, then we may associate a given absorption radius $R$ with our fitted cross-section $\sigma$; that is, $R \equiv \sqrt{\sigma/\pi}$. In order to determine whether our empirical cross sections support such models, we show in Figure~\ref{fig:sigma_v_M_empirical_SiIV} how the integrated cross section varies with stellar mass for the best-fit model as well as a selection of high-probability models. Power-law relationships $R \propto M_*^\beta$ will manifest as straight lines in $\log(\sigma)-\log(M_*)$ space.

The top panel confirms that the cross section for weak absorbers ($\log(N) > 12$) varies roughly as a power-law function of $M*$ for $M_*>10^6\msun$. For lower masses, it declines sharply owing to our $y_\mathrm{max}$ requirement (check \#~\ref{req:ymax} in Section~\ref{ssec:empAbsGal-priors}). For stronger absorbers ($\log(N) > 13$), the behavior is similar, but the cutoff mass increases. 

The bottom panel indicates that power-law relations between absorber size and host mass do not emerge generically from our formalism. With few exceptions, $\sigma$ increases with $M_*$, but the appropriate power-law parameter is itself a function of mass, rising from $\beta\approx0.25$ for weak systems to 1 and even 2 for strong absorbers and low masses. We conclude that the convenient assumption of a Holmberg-like power-law dependence between absorbing region size and galaxy mass may be acceptable for narrow intervals in column density and stellar mass, but it is empirically unsupported.

As an aside, the bottom panel reveals a discontinuity in the derived power-law slope around $10^{7.5}\msun$. This is an artefact of the boundary in the~\citet{sunf16} model between the region where the stellar baryon fraction was empirically constrained versus the region where extrapolation was necessary. It arises here because our model relates the absorption cross section to the host halo's virial radius, which we in turn derive using the~\citet{sunf16} baryon fractions. 

\begin{figure}
\centering
\includegraphics[width=90mm]{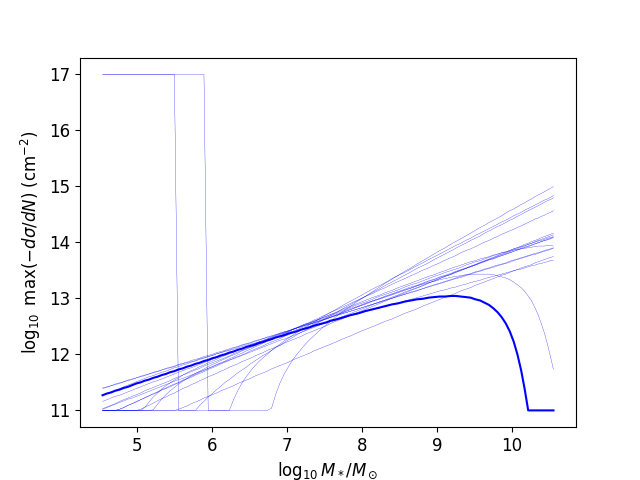}
\caption{The most likely column density $N_\mathrm{max}$ (Equation~\ref{eqn:Nmax}) as a function of stellar mass for observed SiIV absorbers at $z=5.595$. The solid curve indicates the best-fit model from Figure~\ref{fig:dndXdN_emp_SiIV} while the light curves indicate other models randomly drawn from the 67\% confidence region.
}
\label{fig:NMax_Mstar_empirical}
\end{figure}

\subsection{The Most Likely Absorber and the Faint Fraction}\label{ssec:empAbsGal-NMax}

Our model permits galaxies at fixed stellar mass to host absorbers spanning a broad range of strengths and may therefore appear to be at odds with observations that sightlines passing near bright galaxies tend to uncover strong metal absorbers~\citep{stei10}. We address this concern by deriving two complementary observational tests. First, we compute the \emph{most likely} column density as a function of stellar mass. This is simply the column density $N_\mathrm{max}(M_*)$ for which the negative derivative of Equation~\ref{eqn:sigmaFit} with respect to column density is maximized: 
\begin{equation}\label{eqn:Nmax}
N_\mathrm{max}(M_*) = N | -\frac{\mathrm{d}}{\mathrm{d}N}  \sigma(M_*,>N) = \mathrm{MAX} \left[ -\frac{\mathrm{d}}{\mathrm{d}N}  \sigma(M_*,>N) \right]
\end{equation}
Figure~\ref{fig:NMax_Mstar_empirical} shows this quantity for the same high-likelihood models that were displayed in the top-right panel of Figure~\ref{fig:dndXdN_emp_SiIV}. Generally, $N_\mathrm{max}(M_*)$ increases with $M_*$, indicating that strong absorbers remain most likely around massive galaxies. In most cases, $N_\mathrm{max}$ varies roughly as $M_*^{0.5}$, where the power-law dependence reflects the inferred value of $l_1$ dependence (Table~\ref{tab:xFits}). This empirically supported behavior is qualitatively consistent with the simulated cross-sections in Figure~\ref{fig:sigmaN}, which turn over at a higher column density for galaxies with higher stellar masses, although the observationally inferred value for $l_1$ is slightly shallower. The inference that stronger absorbers generically trace more massive galaxies also appears to be statistically robust: parameters $l_0$ and $l_1$ are constrained to fall in a region that is much smaller than the search region, as indicated by the $l_1-l_0$ covariance plot in Figure~\ref{fig:dndXdN_emp_SiIV}. 

\begin{figure}
\centering
\includegraphics[width=90mm]{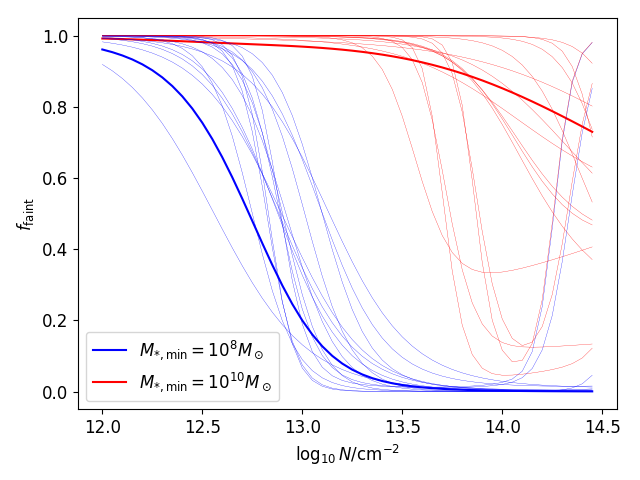}
\caption{The inferred fraction $f_\mathrm{faint}$ of absorbers whose hosts are less massive than a threshold mass as a function of SiIV column density. Two threshold masses are given. Thick and thin curves correspond to the best-fit model from Figure~\ref{fig:dndXdN_emp_SiIV} while the light curves indicate other models randomly drawn from the $1\sigma$ region in the posterior probability distribution. For column densities where the CDD is constrained, $f_\mathrm{faint}$ declines with increasing column density. $f_\mathrm{faint}$ is permitted to rise to high columns, reflecting uncertainty in the stellar masses of galaxies forming in low-mass halos and the lack of constraints on the CDD at high columns.
}
\label{fig:fUn}
\end{figure}

Figure~\ref{fig:NMax_Mstar_empirical} confirms that our empirical model \emph{generally} associates strong SiIV absorbers with massive galaxies and weak absorbers with low-mass ones, qualitatively consistent with observations~\citep{stei10}. It is possible in principle to derive the converse; that is, the most likely \emph{host mass} for a given \emph{column density}. We will consider this in future work. For the present, we instead explore implications of the prediction that a fraction of strong absorbers arise around hosts that are too faint to be observed. This predicted ``faint fraction" $f_\mathrm{faint}$ remains uncertain because, as Figures~\ref{fig:dndXdN_Mlim_emp_SiIV}--\ref{fig:sigma_v_M_empirical_SiIV} confirm, a degeneracy between host mass and geometric cross section remains even though our model is physically motivated. Direct observations of the association between absorbers and their host galaxies are required to break this degeneracy. In practice, however, surveys for absorber hosts often turn up null results, which yield only limits on the typical host mass or luminosity~\citep{wu2023,schroetter2021}. In order to demonstrate the constraining power of such null results, we show in Figure~\ref{fig:fUn} how $f_\mathrm{faint}$ varies with column density and detection limit. For convenience, we define $f_\mathrm{faint}$ as the fraction of absorbers stronger than a column density $N$ originating in galaxies whose stellar masses fall below $M_{*,\mathrm{min}}$. This mimics an imaging survey whose completeness is a step function in stellar mass. With this definition, $f_\mathrm{faint}$ is computed as
\begin{align}
f_\mathrm{faint} \equiv \frac{\int_0^{M_{*\mathrm{,min}}} \frac{dn}{dM_*} \sigma(>N|M_*) dM_*} {\int_0^{\infty} \frac{dn}{dM_*} \sigma(>N|M_*) dM_*}
\end{align}
We show in Fig.~\ref{fig:fUn} how $f_\mathrm{faint}$ for SiIV absorbers varies with minimum column density $N$ for $M_{*,\mathrm{min}}=10^8$ and $10^{10}\msun$. For sufficiently weak absorbers, $f_\mathrm{faint}$ approaches unity irrespective of the mass threshold because all models attribute weak absorbers predominantly to faint, low-mass galaxies even though weak absorbers are permitted about massive galaxies (Fig.~\ref{fig:dndXdN_Mlim_emp_SiIV}). $f_\mathrm{faint}$ then declines with increasing $N$. While this association is to some extent enforced by our requirement that the model forbids unphysical total ion yields (requirement~\ref{req:ymax} above), it is also natural consequence of the fact that bright galaxies and strong absorbers are both rare. Finally, $f_\mathrm{faint}$ increases with $M_{*\mathrm{,min}}$, reflecting the need for deep followup surveys to identify absorber hosts with high completion.

Two examples of nontrivial behavior in Figures~\ref{fig:NMax_Mstar_empirical}--\ref{fig:fUn} merit discussion. First, some models envision strong absorbers as compact systems around low-mass galaxies, resulting in an upturn in $f_\mathrm{faint}$ for absorbers with $\log(N) > 14$. This reminder of the degeneracy between geometric cross section and host mass is not obviously unphysical because requirement~\ref{req:ymax} prevents the total ion abundance from exceeding a reasonable stellar metal yield. Additionally, not all strong absorbers at $z>5$ are associated with observable hosts~\citep{wu2023}. Nonetheless, a model that predicts abundant strong absorbers hosted entirely by dwarf galaxies would be unexpected given the robust association between strong absorbers and bright galaxies at lower redshifts~\citep{stei10}. This prediction can be tamed by enforcing upper limits on the predicted abundance of strong systems and by incorporating results from surveys for hosts of strong absorbers. Second, the typical $N_\mathrm{max}(M_*)$ for massive galaxies declines in two of the models that we show in Figure~\ref{fig:NMax_Mstar_empirical}. Such models preferentially surround massive galaxies with weak absorbers, once again in conflict with results at lower redshifts. This could likewise be addressed through statistical constraints on the absorber-galaxy relationship.

\section{Discussion}\label{sec:discuss}
Our motivation was to ask whether the CDD, when combined with constraints on the underlying SMF, reveals the galaxy mass that dominated star formation during the hydrogen reionization epoch. This leverages the idea that metal absorbers trace recent star formation more completely than rest-frame ultraviolet/optical light, effectively probing to fainter luminosities and lower masses even than JWST imaging surveys. To this end, we have derived a new six-parameter analytical framework for associating galaxies with metal absorbers that reconciles the galaxy stellar mass function with the absorber CDD (Figure~\ref{fig:dndXdN_emp_SiIV}). Based on Equation~\ref{eqn:dndXdN}, it can trivially be modified to quantify absorber strength using equivalent width rather than column density, galaxy luminosity in place of stellar mass, or to consider other ions than SiIV. It generalizes the relationship between galaxy mass and cumulative absorption cross section that arises naturally in cosmological simulations. It is significantly more flexible than analyses emerging from Equation~\ref{eqn:dndX}. It relaxes the need for assuming a particular dependence of cross section on luminosity although traditional power-law dependencies are not excluded (Figure~\ref{fig:sigma_v_M_empirical_SiIV}). It naturally permits galaxies at fixed stellar mass or luminosity to host absorbers with a wide range of strengths, in agreement with observational results~\citep{stei10,bord24}. 

As a probe of star formation in low-mass halos whose galaxies cannot be observed in emission, the method may be used in two complementary ways reflecting the lingering degeneracy between galaxy abundance and cross section that emerges from Equation~\ref{eqn:dndXdN} and is visible in Fig.~\ref{fig:dndXdN_Mlim_emp_SiIV}. First, one may simply extrapolate an adopted SMF to lower masses and constrain the range of CACS that can account for the CDD. This is equivalent to assuming that the dependence of stellar mass on halo mass is known \emph{a priori}. The model may assign a wide range of CACS to galaxies at fixed stellar mass subject to the condition that the total mass does not yield the maximum ion yield (Condition~\ref{req:ymax} in Section~\ref{sec:empAbsGal}). This has been our approach. An alternative approach is to specify the CACS \emph{a priori} and use it to constrain the stellar mass function below current detection limits. In this case, the inferred low-mass stellar mass function constrains the star-formation efficiency in low-mass halos. We leave an exploration of this alternative approach to future work but note several ways to extend our current work. 

First, the results in Figure~\ref{fig:dndXdN_emp_SiIV} indicate degeneracies between each normalization-slope parameter pair. The anti-correlations indicate that models are preferred in which galaxies within a dominant mass range have a particular CACS while the absorbing properties of galaxies outside this mass range are less well-constrained. Such a mass range dominance is expected if (low-mass, high-mass) galaxies cannot contribute many absorbers owing to their low (metal productions, cosmic abundances). Indeed, Figures~\ref{fig:dndXdN_emp_SiIV}--\ref{fig:dndXdN_Mlim_emp_SiIV} suggest that the most well-constrained SiIV absorbers are those with $\log(N) =$12.5--13.5, which are in turn hosted by galaxies with $\log(M_*/\msun) = $7--9. Improved CDD measurements spanning a wider range of column densities will alleviate these degeneracies.

Concurrently, Figure~\ref{fig:fstar_mhalo} shows that the expected stellar baryon fractions in low-mass halos at $z=6$ remain uncertain at the $\approx$0.5 dex level. Assuming overall lower stellar baryon fractions would lead either to a larger extravirial contribution, a lower minimum galaxy mass, or both. Improved observational constraints on the overall abundance of low-mass galaxies will provide key constraints on theoretical models of low-mass galaxy formation. State-of-the-art surveys now constrain the abundance and clustering of galaxies with stellar mass $>10^{8.3}\msun$ at $z>5$~\citep{stefanon21,weib2024,paquereau2025}. At $z=6$, such galaxies probably inhabit halos more massive than $\log(M_h/\msun) = 10.5$~\citep{paquereau2025}. However, models and observations indicate that star formation probably occurs in halos down to the HI cooling threshold of $\log(M_h/\msun) = 8.2$~\citep{nebrin23,dhandha2025}. Na\"ively extrapolating two orders of magnitude down from observed systems indicates expected stellar masses of $\sim3\times10^4\msun$. In other words, current surveys could still be four orders of magnitude too shallow to identify the most abundant galaxy population. 

These considerations reinforce the need for continued theoretical inquiry into ``reasonable" extrapolations of the stellar baryon fraction to low masses. Our baseline empirical model~\citep{sunf16} offers a broad range of extrapolations, which in turn leads to substantial uncertainty in the abundance of low-mass galaxies. Our simulation includes star formation in the extrapolated range (Figure~\ref{fig:fstar_mhalo}), but its dynamic range does not include the HI cooling limit. With a star particle mass of $m_*=1.28\times10^5\msun$, it resolves star formation only in galaxies more massive than $\approx 64m_* = 8.2\times10^6\msun$. Complementary efforts with substantially higher mass resolution will establish the expected abundance of lower-mass galaxies~\citep[for example,][]{feldmann2023,kannan2025}. Such models will yield first-principles predictions regarding the geometric cross section to absorption at low masses. Additionally, the predicted overall abundance of galaxies below current observational limits will reduce the modeling uncertainty: if the stellar baryon fraction in low-mass halos is (weaker, stronger) than we assume, then the inferred geometric cross-sections are correspondingly (larger, smaller).

Another observational constraint that we have neglected is the abundance of ultra-strong absorbers. Figure~\ref{fig:dndXdN_emp_SiIV} shows that the predicted abundance of absorbers stronger than the strongest observed system span a broad range. As the observational completeness for such systems probably approaches unity, it should be straightforward to derive upper limits on their abundance and exclude models that violate them. 

Direct observational constraints on the absorber-host relationship will collapse the model space in two complementary ways. Figure~\ref{fig:fUn} shows that acceptable models make a broad variety of predictions regarding the fraction of absorbers that do not have an observable host galaxy, hence even followup imaging surveys that return null results can be used to constrain the model's parameter space, improving the predictive power of extrapolations to lower masses. At the same time, the inverse of the faint host fraction, the ``covering fraction," defined as the fraction $f_c$ of galaxies for which absorption is observed, can readily be derived by combining spectroscopic and imaging or multi-object spectrograph campaigns~\citep[for example,]{chen2001,bordoloi2014,burchett2016,schroetter2021}.  The relative abundance of observational constraints on $f_c$ thus represents an obvious test of any model for the galaxy-absorber relationship. However, making contact with these observations requires the introduction of at least one additional parameter in order to determine the distance out to which absorption is observed, and in any case such observations remain currently unavailable for SiIV absorbers in the reionization epoch, hence we defer further consideration of $f_c$.

\section{Summary}\label{sec:sum}
We have introduced a new technique for studying the galaxy-absorber relationship by using the observed CDD and SMF to infer the CACS. This method incorporates significantly more information than the integrated absorber line incidence, allowing a more natural mapping between stellar mass and column density. In order to implement this technique, we have used cosmological simulations to derive a parameterized model for the CACS as a function of stellar mass and verified that it contains enough information to reconstruct the full CDD. Applying it to observations, we find the following results:
\begin{enumerate}
\item A direct comparison between the observed SiIV and virial line incidences at $z\sim$5--6 indicates that the observations trace virialized gas in hydrogen-cooling halos.
\item Our generalized CACS model constrains the unobserved host galaxies of weak absorbers by connecting their spatial abundance and geometric cross section. By assuming one of these, the other is obtained in a way that accounts naturally for the full SMF and CDD.
\item The parameterized CACS readily reconciles the observed CDD with the observed galaxy SMF, but the extrapolated CDD is uncertain at both the strong and the weak ends.
\item The resulting parameter space contains lingering degeneracies. To some extent, this is an inevitable consequence of constraining six parameters using four measurements. However, the inherent degeneracy between geometric cross section and galaxy number density remains even though we invoke a physically motivated CACS model. Future observations spanning a broader range in column density will probe the absorbing properties of a broader range of galaxy masses, weakening the degeneracies.
\item Observations suggest an association between strong absorbers and massive galaxies, driven in part by the requirement that the galaxy ion yield respect a physically motivated maximum.
\item Observations likewise suggest a stronger association between weak absorbers and faint galaxies. Indeed, the fraction of $z\gtrsim6$ SiIV absorbers whose host galaxies are less massive than $10^8\msun$ only drops below 50\% for column densities larger than $10^{12.7}$cm$^{-2}$ (Figure~\ref{fig:fUn}), emphasizing the complementarity in emission versus absorption selections.
\item When interpreted using our generalized CACS model, observations do not require the adoption of a universal power-law relation between stellar mass and the size of the absorption region.
\item Future follow-up surveys quantifying the fraction of absobers for which hosts are detected will add complementary constraints to our modeling framework. Indeed, even a precise measurement of the fraction of \emph{strong} absorbers $\log(N_\mathrm{SiIV}) > 14$ whose hosts are undetected would already constrain the model (Figure~\ref{fig:fUn}).
\end{enumerate}

\appendix
\section{Cosmology Conversions}
We here describe how we convert from the published SMFs to our adopted cosmology and initial mass function (IMF). To convert from either the Chabrier~\citep{chab03} or the Kroupa~\citep{krou01} IMF to the Salpeter~\citep{salpeter1955} IMF, we use the mass-to-light ratio conversions $\kappa_\mathrm{chab}=0.63$, $\kappa_\mathrm{kroup}=0.67$~\citep{maddick14}. Our simulation uses the Kroupa IMF interchangeably with the ``135\_100" IMF from BPASS version 2.2~\citep{eldr17}, and we adopt a Hubble parameter $h=0.6774$. For comparison with other works, we define $h7\equiv0.7$ and invoke the rule that absolute luminosities and hence stellar masses vary as $h^{-2}$ while comoving volumes vary as $h^{-3}$.

\citet{weav22} assume the Chabrier IMF and $h=h7$. We convert their published SMFs as follows:
\begin{eqnarray*}
M_*  & \rightarrow  & (\kappa_\mathrm{kroup} / \kappa_\mathrm{chab}) * (h/h7)^{-2} M_* \\
\Phi  & \rightarrow & (h/h7)^3 \Phi \\
\end{eqnarray*}

\citet{stefanon21} use the Salpeter IMF with $h=h7$. We convert their published SMFs as follows:
\begin{eqnarray*}
M_*  & \rightarrow  & \kappa_\mathrm{kroup} (h/h7)^{-2} M_* \\
\Phi & \rightarrow & (h/h7)^3 \Phi \\
\end{eqnarray*}

\citet{weib2024} use the Kroupa IMF with $h=h7$. We convert their published SMFs as follows:
\begin{eqnarray*}
M_*  & \rightarrow  & (h/h7)^{-2} M_* \\
\Phi & \rightarrow & (h/h7)^3 \Phi \\
\end{eqnarray*}

\citet{sunf16} use a Salpeter IMF with $(\Omega_b, \Omega_M) = (0.046, 0.28)$, which differs slightly from our
adopted values of (0.0486, 0.3089). We therefore convert their published stellar baryon fractions as follows:
\begin{eqnarray*}
f & \rightarrow & \kappa_\mathrm{kroup} \frac{\frac{0.046}{0.28}}{\frac{0.0486}{0.3089}} f
\end{eqnarray*}

\section{Corrected Fit Parameters}
The published fit parameters for the mass- and redshift-dependent star formation efficencies in~\citet{sunf16} do not contain enough significant figures to reproduce their published results. We obtained the fit parameters including more significant figures from the lead author by private communication and include them here for reference: 
\begin{eqnarray*}
a0 & = & -178.16675545815485 \\
a1 & = & 39.11080270601153\\
a2 & = & 10.736952692144744\\
a3 & = & -0.9171676971275342\\
a4 & = & -2.8737296197964675\\
a5 & = & 0.07084577245777937
\end{eqnarray*}

\section*{Author Contributions}
K.F.\ led all aspects of the analysis and writing although S.P.\ and N.N.\ contributed to early development of the maximum-likelihood analysis. A.C., E.H., S.K., and F.H.\ contributed equally to interpretation and presentation.

\section*{Acknowledgements}
KF thanks John Weaver for clarifications regarding the~\citet{weav22} SMFs; Jason Sun for forwarding corrected fit parameters from the~\citet{sunf16} stellar mass functions; Brian O'Shea and Moire Prescott for advice and encouragement; and Joseph Burchett for thoughtful feedback on an initial draft. Dark Matter halo mass functions are computed using Simeon Bird's publicly-available halo\_mass\_function.py package. We thank the anonymous referee for many thoughtful suggestions that improved the draft. This work utilized resources from the New Mexico State University High Performance Computing Group, which is directly supported by the National Science Foundation (OAC-2019000), the Student Technology Advisory Committee, and New Mexico State University, and benefits from inclusion in various grants (DoD ARO-W911NF1810454; NSF EPSCoR OIA-1757207; Partnership for the Advancement of Cancer Research, supported in part by NCI grants U54 CA132383 (NMSU)). KF gratefully acknowledges support from the National Science Foundation under Award Number 2006550. KF additionally acknowledges support from NASA through program \#JWST-AR-05779.001-A, which was provided by NASA through a grant from the Space Telescope Science Institute, which is operated by the Association of Universities for Research in Astronomy, Inc., under NASA contract NAS 5-03127. This research has been enabled by the NASA Astrophysics Data System and the arXiv eprint service. The Cosmic Dawn Center of Excellence is funded by the Danish National Research Foundation under then Grant DNRF140.



\bibliographystyle{mnras}
\bibliography{absMF} 


\end{document}

%% file: mymacros_nojournals
\newcommand{\K}{{\rm K}}

\newcommand{\hmpc}{h^{-1}{\rm Mpc}}

\newcommand{\msun}{M_{\odot}}

\newcommand{\s}{{\rm s}}